\documentstyle[psfig]{caosp}
\def\v{{$v\sin i$}\ }

\begin{document}
\pubyear{1993}
\volume{23}
\firstpage{55}
\htitle{Why are magnetic Ap stars ...}
\hauthor{K. St\c epie\'n}
\title{Why are magnetic Ap stars slowly rotating?}
\author{K. St\c epie\'n \inst{1}}
\institute{Warsaw University Observatory, Al. Ujazdowskie 4, 00-478
Warszawa, Poland}

\maketitle
\begin{abstract}
Observational data on rotation of Ap stars suggest that the bulk of
their rotation rates
form a separate Maxwellian distribution with an average value 3-4 times
lower than the normal star distribution. No evidences for a significant
angular momentum (AM) loss on the main sequence (MS) have been found.
It is thus concluded that Ap stars must lose a large fraction of their
initial angular momentum (AM) in the pre-MS phase of evolution, most
probably as a result of the interaction of their primordial magnetic
fields with accretion disks and stellar winds. The
observationally most acceptable values of accretion rate from the disk,
$10^{-8}  M_{\sun}$ /year, of mass loss rate via a magnetized wind, $10^{-8}
M_{\sun}$ /year, and of the surface magnetic field, 1 kG on the ZAMS,
result in the AM loss in full agreement with observations. 

There exists a separate group of extremely slowly rotating Ap stars, with
periods of the order of 10-100 years. They are too
numerous to come from the distribution describing the bulk of Ap stars. It
is conjectured that their extremely low rotation rates are the result of
additional AM loss on the MS. 
\end{abstract}

\section{Introduction}
\label{intr}

The A and B type stars with peculiar spectra have been recognized as
having unusually sharp spectral lines as soon as their spectra were analyzed 
with large enough dispersion. In fact, low \v values were the
reason why Babcock (1947) included them into his search for stellar
magnetic fields. He believed at that time that strong magnetic fields
are coupled with rapid rotation but because of instrumental reasons he
selected early type stars with sharp lines, assuming that the low values of
\v are the result of low $i$, not $v$ values. Later, when the variation
periods of Ap stars (typically of several days) were identified as
their rotation periods,  according to the oblique rotator theory
(Stibbs, 1950), it became apparent that the low \v values are the result
of the low rotation rates
of these stars. The presently
known values of the rotation periods of peculiar stars
are in agreement with the conclusion that the stars rotate on average much
slower that normal stars of the same spectral types (Catalano \&
Renson, 1997 and references therein).

The observed chemical peculiarities of Ap stars are explained as resulting
from the diffusion of different ions under the influence of the radiative
force (Michaud, 1970; Michaud \& Proffitt, 1993).
A slow rotation rate was assumed to be {\it condicio sine qua
non} for a diffusive segregation of elements to occur in atmospheres of
these stars although the role of the magnetic field must also be very
important for the development and distribution of over- and
under-abundances over the stellar surface (Pyper, 1969; Babel 1993).
Recently Abt \& Morrel (1995) went one step further suggesting that the
slow rotation is also a sufficient condition for the Ap star phenomenon to
occur. They propose the
existence of a threshold for A stars, such that all stars
rotating slower than that should be chemically peculiar, even if not yet
recognized as such.

\begin{table}[t]
\small
\begin{center}
\caption{Ap stars with extremely long periods.}
\label{t1}
\begin{tabular}{llll}
\hline\hline
HD & Sp & $P$ (years)\\
\hline
9996 & A0 SiCrSr & 22\\
94660& A0 EuCrSi & $\ga$ 7.5\\
110066& A1 SrCrEu& 13.5 or 27\\
137949& F0 SrCrEu& $\ga$ 75\\
187474& A0 EuCrSi& 6.5\\
201601& A9 SrEu &  75\\
\hline\hline
\end{tabular}
\end{center}
\end{table}

The rotation distribution of Ap stars cannot be interpreted as a slow rotation
tail of the normal star distribution. A statistical investigation of rotation
velocities of peculiar stars indicates that their distribution can be
approximated by the Maxwellian one with an average value 3-4 times lower
that for normal stars (Preston, 1970; Wolff, 1981; Abt \& Morrel, 1995).
In addition, several Ap stars with extremely long rotation periods of
more than 5 years are known (Table 1). Note that all belong to cooler
Ap stars of the CrSrEu type.
The periods of these stars, if interpreted as rotation periods,  
cannot be part of the Maxwellian distribution describing the bulk of
the rotation periods of peculiar stars, because the probability of finding
even one rotator with such a long period in the known sample of Ap stars is
exceedingly small (Preston, 1970). We have to assume that they form a
separate population of Ap stars in which a special mechanism of spin
down is (or has been) operating.

The slow rotation of A and B peculiar stars can be a result of one or
more of the following circumstances:
(i) they are formed from protostellar clouds with particularly low
angular momentum (AM),
(ii) they lose extra AM in the pre main sequence (PMS) phase of
evolution,
(iii) they lose AM on the MS.

The first possibility seems unlikely. Open clusters contain a
substantial number of peculiar stars which have been formed
simultaneously with the other cluster members. Had they been formed from a
low rotation tail of the protostellar cloud distribution, they would have
become part of the same stellar distribution, which contradicts the
observations. 

The second and third possibility were a subject of
investigation of several authors in the past. Abt (1979) measured \v
values of Ap stars in a number of stellar clusters of known ages and
found a significant decrease of the rotation rate of at least hot peculiar
stars with age. That was confirmed by Wolff (1981). Unfortunately, their
conclusion relied heavily on the data points from the Orion association
for which only three \v values had been measured. Wolff (1981) discussed
also possible mechanisms for AM loss of Ap stars, based on the
interaction of their magnetic fields with the stellar environment.
More accurate and numerous data on the rotation periods of
peculiar stars in young clusters obtained later, showed that the period
distribution of peculiar cluster members is indistinguishable from the
distribution of field Ap stars, assumed to be much older (Borra et al. 1985;
North, 1984a, 1987). North (1984b) discussed the dependence of the observed
period of field Ap stars on gravity, treated as a measure of age for a
given mass. He concluded that the rotation period increases when $\log
g$ decreases just as expected from the conservation of AM during the MS
life, without ``...the least suggestion of any braking mechanism''. But
recently Pyper et al. (1998) showed that the short-period Ap star
CU Vir abruptly increased its period by about $5 \times 10^{-5}$ of its
value. Because the reason for this change seems at present completely
obscure, and it is not clear what is its relation to possible
evolutionary period changes during the MS life, that case will be ignored
and it will be assumed in the
following that the observational data do not show significant AM loss
of Ap stars during their MS life.

This leaves us with the hypothesis that progenitors of peculiar stars are
born with normal rotation rates but they lose a large fraction of their
initial AM in the PMS phase, hence they rotate 3-4 times slower than
normal stars when they land on the ZAMS. This hypothesis will be the
subject of the rest of the paper.

\section{PMS evolution of intermediate mass stars}

An important difference appears in the PMS evolution of intermediate
mass stars (IMS), compared to low mass stars. Time spent in a fully
convective phase decreases rapidly for stars with masses above 1.5 $M_{\sun}$,
reaching $10^4$ years for 2 $M_{\sun}$ and zero for masses above 
2.4 $M_{\sun}$ (Palla \& Stahler, 1993). It makes the survival of a
primordial magnetic field much more probable for IMS
than for solar type stars. On the other hand, a convection zone connected
with the deuterium burning is very shallow in IMS, which makes the
existence of strong dynamo generated fields very unlikely. Indeed, the
observations of Herbig Ae/Be stars indicate that their activity level is
not correlated with rotation, as would dynamo theory predict, but
with effective temperature (B\" ohm \& Catala, 1995). The absence
of the dynamo generated fields in IMS is also in agreement with
the observational data on rotation of these stars. The
comparison of the rotation velocities of Herbig Ae/Be stars with the
ZAMS stars of the same mass shows that the IMS do
not lose a measurable amount of AM during the PMS phase if their AM is
conserved in shells during the approach to the ZAMS (B\" ohm \& Catala,
1995). This is not, however, the case for Ap stars which rotate
much slower than the normal stars of the same spectral types
(Wolff, 1981; Abt \& Morrel, 1995).
Consequently, they must lose a large fraction of the initial AM in the
PMS phase.

Detailed models of the PMS evolution of IMS were
computed by Palla \& Stahler (1993). The results show that the
PMS phase of these stars is rather short:
from slightly less than about $10^7$ years for a 2 $M_{\sun}$ down to
$2\times 10^5$ years for a 5 $M_{\sun}$ star. This is substantially
less than adopted e. g. by Wolff (1981) after
Iben (1965). Shorter time scales require a more efficient spin down
mechanism. 

Observations of Herbig Ae/Be stars show the presence of stellar winds
with a mass loss rate of $10^{-8}  M_{\sun}$ /year or more, as well as
the presence of circumstellar matter, very likely in the form of
accretion disks (Catala, 1989; Palla, 1991). We can expect accretion
rates not much different from those observed in T Tauri stars, i. e.
$10^{-9} - 10^{-8}  M_{\sun}$ /year (Basri \& Bertout, 1989).

Considering the AM loss of Ap stars it will be assumed that they
preserve primordial magnetic fields through the protostellar phase and
the magnetic field interacts with both the stellar wind and the accretion
disk during the PMS phase of evolution, which influences the stellar AM.
Details of this process will be discussed in the next Section.

\section{AM loss mechanism of magnetic Ap stars}

We will consider now the evolution of the stellar AM during the PMS
evolution. The AM of a rigidly rotating star is given by $I\omega$, where $I$
is the moment of inertia of the star and $\omega$ its angular velocity.
Assuming that the time derivative of AM is equal to the total torque $T$
exerted on the star we have
\begin{equation}\label{r1}
{{{\rm d}\omega}\over {{\rm d}t}} = {1\over I}\left(T - \omega {{{\rm d}I}\over
{{\rm d}t}}\right).
\end{equation}

According to our assumptions, the total torque will consist of three
parts, $T = T_{mag} + T_{acc} + T_{wind}$, where $T_{mag}$ comes from the
magnetic star-disk linkage, $T_{acc}$ is due to magnetic accretion of the
matter from the disk and $T_{wind}$ is connected with the magnetized wind.
Let us discuss each of these terms separately.

The recent observations indicate that some of Herbig Ae/Be stars are
surrounded by massive, optically thick disks, whereas others are disk-less
(Hillenbrand et al., 1992; Grinin, 1992; B\" ohm \& Catala, 1995;
Corcoran \& Ray, 1997). This suggests that a typical time scale of the disk
life is probably shorter than the PMS life time of an IMS. 

The expression for the magnetic torque was derived by Armitage \& Clarke
(1996). The maximum efficiency of the torque is reached when the radius of
the magnetosphere (identical with the radius of the inner edge of the disk)
is equal to the corotation radius. We will assume this for simplicity, and any
possible variations of efficiency of this or the other considered
mechanisms will be accounted for later by introducing arbitrary, 
multiplicative weights. We have thus
\begin{equation}\label{r2}
T_{mag} = -{{{\mu}^2{\omega}^2}/{3GM}},
\end{equation}
where $\mu = BR^3$ is the stellar magnetic dipole moment, assumed here to
be constant, $G$ is the gravity constant, $B$ is the intensity of the
surface magnetic field, and $R$ and $M$ are radius and
mass of the star, respectively.

To consider $T_{acc}$ it is assumed that the matter is accreted from the
inner edge of the disk along the magnetic field lines. If the radius of
the inner edge is much larger than $R$, the accretion torque can be
approximated by
\begin{equation}\label{r3}
T_{acc} = {\dot M_{acc}{(GM)}^{2/3}/{\omega}^{1/3}},
\end{equation}
where ${\dot M_{acc}}$ is the accretion rate.

The expression for the torque due to a magnetized wind in case of a dipolar
magnetic field is given by (St\c epie\'n, 1995)
\begin{equation}\label{r4}
T_{wind} = - {\omega\over 3}{\dot M}^{3/5}_{wind}
R^{3/5}\mu^{4/5}(2GM)^{-1/5},
\end{equation}
where ${\dot M}_{wind}$ is the mass loss rate via the magnetized wind.

The equation (1) for the angular velocity evolution assumes now the form
\begin{equation}\label{r5}
{{{\rm d}\omega}\over {{\rm d}t}} = {1\over I}\left[{\dot
M_{acc}}{{(GM)}^{2/3}\over {\omega}^{1/3}} - {{{\mu}^2{\omega}^2}\over
{3GM}} -  {\omega\over 3}{\dot M}^{3/5}_{wind}
R^{3/5}\mu^{4/5}(2GM)^{-1/5} -  \omega {{{\rm d}I}\over
{{\rm d}t}}\right]
\end{equation}

Equation (5) is the basic equation solved numerically for the
adopted values of free parameters. The discussion of free parameters and
the results are given in the next Section.

\section{Results and discussion}

The calculations have been carried out for two values of the stellar mass,
2 and 3 $M_{\sun}$. The time scales of the PMS evolution, and the
dependence of the moment of inertia and stellar radius on time, i.e. $I(t)$
and $R(t)$, were taken from models computed by Palla \& Stahler (1993)
(see also B\" ohm \& Catala, 1995).  Based on the observational results
about the accretion rate and mass loss via winds of Herbig Ae/Be stars
given in Section 1, the following values have been adopted as typical: 
${\dot M}_{acc} = 10^{-8} M_{\sun}/{\rm year}$
and ${\dot M}_{wind} = 10^{-8} M_{\sun}/{\rm year}$. The value of the magnetic moment
was adopted as $\mu = 2.7\times 10^{36}$ in cgs units, which corresponds to
the 1 kG dipole field on a 2 $R_{\sun}$ star. To allow for a possible
variation of these values as well as other factors modifying the efficiency
of all the considered mechanisms, arbitrary weights were added to the
the first three terms in equation (5). 

Figure 1 (upper) compares the variation of the rotation period of a 2
$M_{\sun}$ star when its AM is preserved during the PMS evolution (solid
line) and when the consecutive AM change mechanisms, described by the first
three terms in brackets on the right hand side of equation (5) are added.
It is assumed that the star emerges from the protostellar phase after
$10^6$ years with a rotation period of 5 days, and the PMS phase ends
after $8\times 10^6$ years (Palla \& Stahler, 1993). When AM is conserved
in a rigidly rotating star its rotation period decreases down to a value of
0.55 of a day on the ZAMS (solid line). When only the accretion is added,
the ZAMS period is even shorter, and it is equal to 0.15 of a day (dotted
line), because the accretion of a high AM matter from the disk increases
the stellar AM, hence spins up the star. When only a wind is added, the
resulting ZAMS period reaches a value of 1.35 day (dotted-broken line).
The most powerful mechanism influencing the
rotation period of a PMS star is the interaction of its magnetic field with
the disk. If the disk is massive enough, it will force the stellar rotation
in a relatively short time to an approximate value of the corotation
period at the edge of the magnetosphere (Armitage \& Clarke, 1996). In our
case this is close to 5 days, hence the rotation period of the considered
star stays close to this value through the whole PMS evolution (broken
line). 
For the 
initial rotation periods shorter than 5 days the field-disk linkage slows
down the rotation but for values longer that that it spins up the star, so
that the final value of about 5 days is always reached.
\begin{figure}[t]                                                           
\psfig{figure=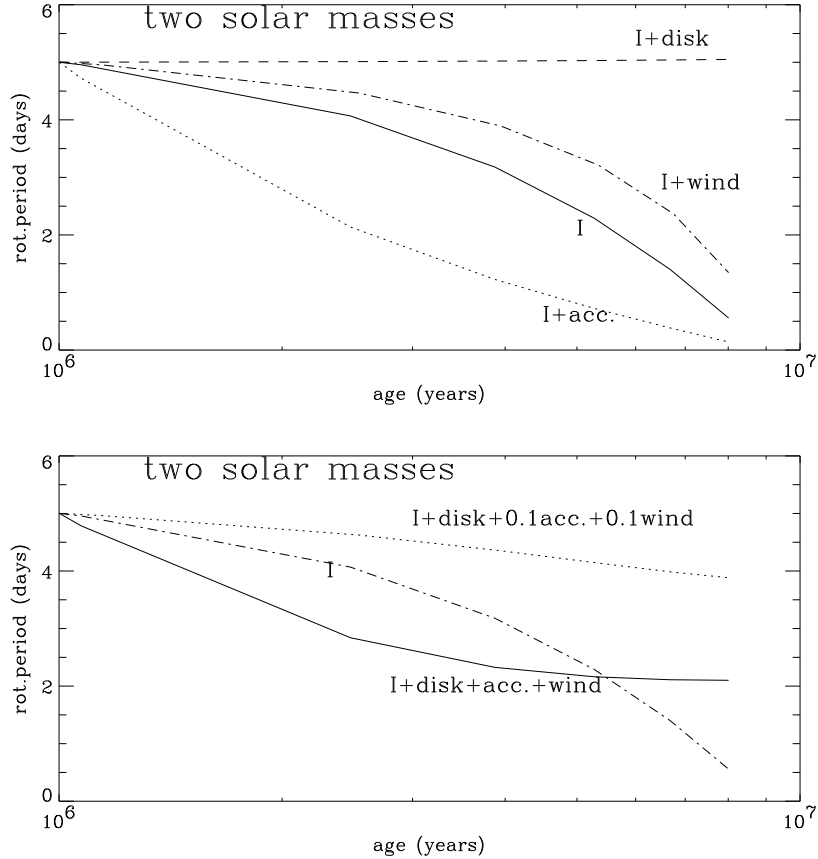,height=12cm}                                        
\caption{The results of model calculations for a 2 $M_{\sun}$ and
different mechanisms for stellar AM change (see text)}
\label{fp}
\end{figure}

Figure 1 (lower) demonstrates the result of a simultaneous action of
all the considered mechanisms. The constant AM case is repeated from the
upper part of the figure for comparison as a dotted-broken line. 
The solid line describes the evolution of the rotation
period of the considered star when all terms in equation (5) are taken into
account with weights equal to 1. The resulting ZAMS rotation period is
equal to 2.1 days, about four times longer than in case of constant AM.
This agrees well with the observations. In addition, the evolution of
the rotation period is shown in the case when the importance of the accretion
and wind is decreased by a factor of 10, i. e. the corresponding terms in
equation (5) were taken with weights equal to 0.1. Physically, this
corresponds to e. g. lower accretion and mass loss rates, or shorter time
scales of both phenomena. The resulting ZAMS rotation period is equal now to
3.9 days. This can be compared with the value of 1.25 days expected when
the considered star preserves AM in shells during its approach to the ZAMS.

Similar calculations have been obtained for a 3 $M_{\sun}$ star. The main
difference between the two cases is the time scale of the PMS evolution
which is 7 times shorter for the 3 $M_{\sun}$ star than for the 2
$M_{\sun}$ star. To get a 3-4-fold increase of the rotation period, a
relatively more efficient AM loss is required. This can be achieved e. g. by
an increase of the weight of the term describing the field-disk interaction
up to a value of 2 -- equivalent to the increase of $\mu$ by a factor of
1.4, see equation (5). Note that this corresponds to a surface magnetic
field of about 1.5 kG which would show up in observations as a longitudinal
magnetic field of only about 500 G (Preston, 1971).

One concludes that the presence of a moderate primordial magnetic field can
explain the observed difference between the average rotation rate of Ap and
normal stars. The required values of the parameters describing the interaction
between a star and its environment are within the observed
ranges. Because the time scales of the existence of disks and winds, mass loss
and accretion rates, and the intensity of the magnetic field are expected
to vary randomly from one PMS star to another, the resulting ZAMS rotation
period is not expected to be strongly correlated with any single parameter.

The discussed AM loss mechanism cannot, however, explain the extremely long
rotation periods observed in some stars (Table 1). The required values of
the considered parameters are unreasonable. Therefore, it is suggested that
those stars lose AM also on the MS because of some exceptional circumstances. 
Because the accretion disks are not observed around young MS
stars we reject this mechanism. A continuous mass loss via a magnetized
wind is a more realistic possibility. With the same intensity as adopted here
for the PMS phase, the wind can spin down the star to a rotation period of the
order of 100 years just in $2\times 10^7$ years, which is a tiny fraction
of the total MS lifetime of a star with a mass below 2 $M_{\sun}$. A less
intense wind would need, of course, a correspondingly longer time.

\acknowledgements
This work was partly supported by the grant 2 P 03D 010 12 from the
Committee for Scientific Research.

\end{document}